# Physics Informed Optimal Homotopy Analysis Method (PI-OHAM): A Hybrid Analytical–Computational Framework for Solving non-linear Differential Equations

Ziya Uddin[1,2]
[1]*SoET, BML Munjal University, Gurugram, Haryana, India*
[2]*Center for Advanced Data and Computational Science, BML Munjal University, Gurugram, Haryana, India*
**\* Corresponding author:** ziya.uddin@bmu.edu.in, ziyauddin1982@gmail.com

**Highlights**

- Suggests a new Physics-Informed Optimal Homotopy Analysis Method (PI-OHAM) for the solution of nonlinear differential equations.
- Presents a residual based optimization scheme to tune systematically the convergence-control parameters of HAM.
- Illustrates the technique on the Blasius, a classical nonlinear viscous flow equation, boundary-layer flow equation.
- Demonstrates that PI-OHAM converges more rapidly and is more accurate than conventional HAM and PINNs.
- Bridges the gap between series-based semi-analytical methods physics-informed machine learning, providing interpretable and efficient solutions to nonlinear differential equations.

**ABSTRACT**

We present the Physics-Informed Optimal Homotopy Analysis Method (PI-OHAM) for solving nonlinear differential equations. PI-OHAM, based on classical HAM, employs a physics-informed residual loss to optimize convergence-control parameters systematically by combining data, boundary conditions, and governing equations in the manner similar to Physics Informed Neural Networks (PINNs). The combination of the flexibility of PINNs and the analytical transparency of HAM provides the approach with high numerical stability, rapid convergence, and high consistency with traditional numerical solutions. PI-OHAM has superior accuracy-time trade-offs and faster and more accurate convergence than standard HAM and PINNs when applied to the Blasius boundary-layer problem. It is also very close to numerical standards available in the literature. PI-OHAM ensures analytical transparency and interpretability by series-based solutions, unlike purely data-driven or data-free PINNs. Significant contributions are a conceptual bridge between decades of homotopy-based analysis and modern physics-inspired methods, and a numerically aided but analytically interpretable solver of nonlinear differential equations. PI-OHAM appears as a computationally efficient, accurate and understandable alternative to nonlinear fluid flow, heat transfer and other industrial problems in cases where robustness and interpretability are important.

# 1. Introduction:

The mathematical basis of the modeling of complex phenomena in fluid mechanics, heat transfer, viscoelastic flows, nanofluid transport, and a variety of applied physical sciences is based on nonlinear differential equations. These equations often do not have closed-form analytical solutions, forcing researchers to use either computationally expensive numerical simulations or approximate analytical methods. The classical methods of perturbation, which involve the existence of small or perturbation parameters, are frequently inapplicable to highly nonlinear systems that are frequently found in real-world engineering problems. The Homotopy Analysis Method (HAM) (Liao, 1992) has since 1992, when Liao first introduced it in his PhD dissertation at Shanghai Jiao Tong University, become a revolutionary semi-analytical methodology that allows the construction of convergent series solutions without small-parameter assumptions, linearization, or restrictive perturbation assumptions.

Simultaneously, He (1999) introduced the Homotopy Perturbation Method (HPM), a hybrid of homotopy and classical perturbation methods that produces approximate analytical solutions. Due to its simplicity and ease of application, HPM has been extensively used to nonlinear problems in engineering and applied sciences. HPM, however, does not have a clear convergence-control mechanism, as does HAM, and this restrict its ability to be robust to strongly nonlinear systems.

Sajid and Hayat (2008) conducted a comparative evaluation of HAM and HPM in the framework of nonlinear heat conduction and convection equations. Their analysis showed that HPM is computationally efficient and simple, but HAM is more accurate and convergent reliable, especially in problems that are highly nonlinear. This analogy indicated the significance of convergence-control parameters in homotopy-based methods.

In order to improve convergence and accuracy further, Marinca et al. (2008) proposed the Optimal Homotopy Asymptotic Method (OHAM) that calculates optimal convergence parameters by using asymptotic matching. OHAM has been applied successfully to thin film flow problems, and has been shown to be more computationally efficient and analytically tractable. Further developing this concept, Liao (2010a) introduced an ideal homotopy-analysis method to strongly nonlinear differential equations, formalising systematic methods of choosing convergence-control parameters. In a similar study, Liao (2010b) briefly reviewed the use of HAM in fluid mechanics, noting its usefulness in boundary layer flows, non-Newtonian fluids, and heat transfer issues.

Homotopy methods have also been combined with numerical methods in hybrid formulations to enhance the efficiency of solutions. Motsa et al. (2010) introduced Spectral Homotopy Analysis Method (SHAM) which is a combination of HAM and spectral methods to solve nonlinear boundary value problems with high precision. Building on this, Motsa et al. (2017) introduced the multi-domain spectral relaxation method to chaotic systems of ordinary differential equations, which showed better stability and convergence of highly nonlinear and chaotic dynamics.

Additional improvements are predictive and multi-solution strategies. Abbasbandy and Shivanian (2011) proposed the Predictor Homotopy Analysis Method (PHAM) to improve convergence behavior, whereas Vosoughi et al. (2012) showed that it can be used to capture unique and multiple solution branches in nonlinear reactive transport models. These papers highlight the flexibility of homotopy-based methods to nonlinear systems of interest.

Recent studies have been devoted to the combination of optimization methods and homotopy methods to automate the choice of parameters. Yuan and Alam (2016) integrated HAM and particle swarm optimization (PSO) to solve fractional-order differential equations, and the convergence speed and accuracy were greatly improved. Equally, Zhang et al. (2021) suggested a PSO-based homotopy approach to global numerical optimization, which supports the possibility of metaheuristic optimization in homotopy models.

Thermal and transport phenomena have also been studied using homotopy methods. Abbasbandy (2007) used HAM to nonlinear heat radiation equations and proved that it is effective in thermal modelling. Complex reaction-diffusion systems have been solved using variants of HPM, such as those by Khan and Wu (2011) with He polynomials and Swaminathan et al. (2019) in the modelling of MichaelisMenten kinetics in microdisk biosensors. Odibat (2020) has recently suggested a better optimal homotopy analysis algorithm that increases the reliability of convergence to nonlinear differential equations. Also, Filobello-Nino et al. (2021) introduced a hybrid HPM-Pseudo-Spectral Euler Method (PSEM) to solve the Blasius problem, with better accuracy and computational efficiency.

In general, the development of homotopy-based approaches emphasizes the shift of purely analytical models to hybrid, optimized, and computationally efficient ones. The combination of spectral techniques, predictor techniques and metaheuristic optimization has greatly increased the range of homotopy analysis to highly nonlinear, multi-solution and real-world engineering problems. These advances encourage more studies on sophisticated hybrid homotopy methods of complex nonlinear systems. These hybridized methods are significantly more effective than classical HAM in automating $\hbar$-selection (convergence control parameter) and reducing residual error norms, but most of them use simplified, averaged residuals, which do not adequately capture the full physics (PDEs + boundary/initial conditions + data-driven constraints), so they are not accurate in complex, multimodal, or data-coupled nonlinear PDE systems.

Simultaneously with the maturation of HAM, physics-informed machine learning has transformed the solvers of differential equations with Physics-Informed Neural Networks (PINNs). Hidden physics models by Raissi and Karniadakis (2018) introduced the underlying idea of instantiating physical laws into data-driven models, which allows the discovery of nonlinear partial differential equations using sparse and noisy data. Raissi et al. (2019) formalised this paradigm in the PINN framework, in which the neural networks are optimised to minimise a composite loss function that imposes the governing PDEs, boundary and initial conditions, and available observational data in one way.

PINNs have been shown to be flexible to a broad set of forward and inverse problems. Raissi et al. (2020) demonstrated that PINNs are able to predict hidden velocity and pressure fields in fluid flows with only flow visualizations, which proves their usefulness in data-limited conditions. Lu et al. (2021) created DeepXDE, a full library of deep-learning, to support PINNs and similar physics-informed models of differential equations, both fractional and multi-physics. Since then, PINNs have been applied to difficult regimes like high-speed compressible flows (Mao et al., 2020) and conservation laws using conservative PINNs that strictly enforce physical invariants (Jagtap et al., 2020).

Extensive surveys by Karniadakis et al. (2021) and Cuomo et al. (2022) have made physics-informed machine learning a revolutionary paradigm in scientific computing, between numerical analysis, machine learning, and physical modelling. Specialized research also underscored the increasing role of PINNs in fluid mechanics (Cai et al., 2021). Uddin et al. (2023) proposed the wavelets based PINNs to solve nonlinear ordinary and partial differential equations. Nevertheless, classical PINNs are known to have well-known issues, such as slow convergence, spectral bias, loss-term imbalance, and the inability to find multiple solution branches.

In order to address these shortcomings, recent research has suggested hybrid formulations that combine homotopy concepts with PINNs. Huang et al. (2022) proposed HomPINNs, which incorporates homotopy continuation into the PINN framework to learn a set of solutions of nonlinear elliptic PDEs systematically. Other stability and accuracy enhancements have been made using variational and domain-decomposition techniques, including hp-VPINNs (Kharazmi et al., 2021). These developments indicate that the convergence control of homotopy with physics-informed learning structures is a promising direction to robust and accurate solutions of strongly nonlinear, multimodal, and data-coupled ODE/PDE systems, which is the motivation behind the current study.

The key contributions of this research may be summarized as follows:

- An improved Physics-Informed Optimal Homotopy Analysis Method (PI-OHAM) is developed whereby homotopy parameters are optimized by minimizing a physics-informed residual functional which incorporates differential equation, boundary and initial conditions.
- The approach is applied to a numerically assisted HAM framework, allowing it to be applied to differential equations and nontrivial initial/boundary data in which higher-order terms in the symbols are hard to compute.
- Using benchmark tests on the Blasius boundary-layer equation, PI-OHAM is demonstrated to converge more quickly and more accurately than standard HAM, and to offer a series-like approximation that is interpretable.
- PI-OHAM conceptually fills the gap between existing semi-analytical homotopy approaches and physics-informed machine learning: it has the structured analytic feel of HAM and the systematic philosophy of residual minimization of PINNs, and promises a way forward to efficient and interpretable solvers of nonlinear differential equations.

## 2. Brief Review of the Homotopy Analysis Method

The Homotopy Analysis Method (HAM) is a semi-analytical method of solving nonlinear differential equations, proposed by Liao (1992), that does not need the existence of small or large perturbation parameters. The idea of homotopy on which HAM is founded is to build a continuous deformation of an initial approximation of the solution, so as to offer a flexible and strong framework to deal with strong nonlinearities.
Consider a nonlinear operator equation
$$\mathcal{N}[u(x)] = 0.$$

HAM develops the equation of deformation of the zeroth order as
$$(1-p)\,\mathcal{L}[u(x;p) - u_0(x)] = p\,\hbar\,\mathcal{N}[u(x;p)],$$

In which $p \in [0,1]$ is an embedding parameter, $\mathcal{L}$ is an auxiliary linear operator, $u_0(x)$ is an initial guess that satisfies the specified boundary or initial conditions and $\hbar$ is a convergence-control parameter. The embedding parameter $p$ is a continuous deformation of a simple, linear problem to the original nonlinear problem.

The zeroth-order equation of deformation at $p = 0$ is simplified to
$$\mathcal{L}[u(x;0) - u_0(x)] = 0$$
which yields $u(x;0) = u_0(x)$ that is associated with the original guess which satisfies the given boundary or initial conditions.

The deformation equation at $p = 1$ will recover the original nonlinear problem:
$$\mathcal{N}[u(x;1)] = 0$$
implying that $u(x;1) = u(x)$ the solution of the governing equation being this. Therefore, the solution $u(x;p)$ is continuously deformed as $p$ changes between 0 and 1 from the original approximation $u_0(x)$ to the true solution $u(x)$.

The solution $u(x;p)$ is written in the form of a power series in p:
$$u(x;p) = u_0(x) + \sum_{m=1}^{\infty} u_m(x)\, p^m.$$

Differentiating the zeroth-order deformation equation $m$ times with respect to $p$, then dividing by $m!$, and letting $p = 0$, the $m$-th order approximation $u_m(x)$ is obtained as
$$u_m(x) = \chi_m u_{m-1}(x) + \hbar\, \mathcal{L}^{-1}[R_m(x)],$$
where
$$\chi_m = \begin{cases} 0, & m = 1, \\ 1, & m \geq 2, \end{cases}$$
and the nonlinear residual $R_m(x)$ is defined by
$$R_m(x) = \frac{1}{(m-1)!} \frac{\partial^{m-1}}{\partial p^{m-1}} \mathcal{N}[u(x;p)]\, \big|_{p=0}.$$

under homogeneous boundary or initial conditions. The nonlinear problem is then approximated to give the approximate solution as.
$$u(x) \approx u_0(x) + \sum_{m=1}^{M} u_m(x),$$
where $M$ denotes the truncation order. The HAM series convergence is directly regulated by the auxiliary parameter $\hbar$, thus HAM is especially useful in strongly nonlinear problems.

3. **Proposed PI-OHAM framework and implementation**

In this section the Physics-Informed Optimized Homotopy Analysis Method (PI-OHAM) is introduced that combines the analytical framework of HAM with numerical discretization, optimization, and physics-informed loss minimization. The framework is illustrated on the classical Blasius boundary layer equation, but can be applied to nonlinear ODEs/PDEs or a system of non-linear differential equations.

**3.1 Governing Equation and Homotopy Construction**

The Blasius equation is written as
$$f'''(\eta) + \frac{1}{2} f(\eta) f''(\eta) = 0,$$
subject to
$$f(0) = 0,\; f'(0) = 0,\; f'(\infty) = 1.$$

Following HAM, a homotopy is constructed using a third-order linear operator
$$\mathcal{L}[f] = f'''$$
with an initial guess $f_0(\eta)$ chosen to satisfy the boundary conditions:

$$f_0(\eta) = \eta - \frac{1 - e^{-a\eta}}{a},$$

where $a > 0$ is an **auxiliary shape parameter** introduced to improve convergence.

### 3.2 Numerical Discretization of HAM Operators

In contrast to classical HAM, the suggested framework uses numerical differentiation operators. Second-order accurate finite differences are used to approximate first, second and third-order derivatives on a uniform grid. Linear operator $\mathcal{L}$ is implemented numerically with embedded boundary conditions directly into the operator matrix and allows an efficient sparse LU-based inversion. This numerically aided formulation enables HAM to be used on general nonlinear problems without the need to do symbolic calculation.

In order to allow a completely numerical, but physics consistent, implementation of the Homotopy Analysis Method, the proposed PI-OHAM framework discretizes the governing operators over a finite computational domain. This prevents symbolic differentiation and can be readily extended to nonlinear boundary value problems.

#### 3.2.1 Computational Domain and Grid

The semi-infinite physical region $\eta \in [0, \infty)$ is truncated to $[0, \eta_\infty]$, where $\eta_\infty$ is large enough to impose the far-field boundary condition. A uniform grid is introduced:

$$\eta_i = ih, \quad i = 0, 1, \ldots, N - 1,$$

with step size

$$h = \frac{\eta_\infty}{N - 1}.$$

Let

$$\mathbf{f} = [f(\eta_0), f(\eta_1), \ldots, f(\eta_{N-1})]^T$$

represents the discrete form of the solution.

#### 3.2.2 Discrete Differential Operators

Second-order accurate finite differences are used to determine the first and second-order derivatives. $\mathbf{D}_1$ is the first derivative operator which is defined as:

$$(\mathbf{D}_1 \mathbf{f})_i = \begin{cases} \dfrac{-3f_0 + 4f_1 - f_2}{2\Delta\eta}, & i = 0, \\ \dfrac{f_{i+1} - f_{i-1}}{2\Delta\eta}, & 1 \leq i \leq N - 2, \\ \dfrac{3f_{N-1} - 4f_{N-2} + f_{N-3}}{2\Delta\eta}, & i = N - 1. \end{cases}$$

The second derivative operator $\mathbf{D}_2$ is given by

$$(\mathbf{D}_2 \mathbf{f})_i = \begin{cases} \dfrac{2f_0 - 5f_1 + 4f_2 - f_3}{\Delta\eta^2}, & i = 0, \\ \dfrac{f_{i+1} - 2f_i + f_{i-1}}{\Delta\eta^2}, & 1 \leq i \leq N - 2, \\ \dfrac{2f_{N-1} - 5f_{N-2} + 4f_{N-3} - f_{N-4}}{\Delta\eta^2}, & i = N - 1. \end{cases}$$

The operator composition is used to construct the third derivative operator,
$$\mathbf{D}_3 = \mathbf{D}_1 \mathbf{D}_2,$$
ensuring the consistency between the discrete derivatives. These discrete operators are combined to sparse matrices. The third-order derivative needed by the Blasius equation is a matrix product.

### 3.2.3 Discrete Linear Operator

In HAM, an auxiliary operator $\mathcal{L}$ must be a linear invertible operator. In the case of the Blasius problem, it is selected as
$$\mathcal{L}[f] = f''',$$
Its discrete analog is thus defined as $\mathcal{L} = \mathbf{D}_3$
In order to incorporate the boundary conditions directly into the operator the following changes are made:
$$f(0) = 0 \Rightarrow (\mathcal{L}\mathbf{f})_0 = f_0, \quad and \ f'(0) = 0 \Rightarrow (\mathcal{L}\mathbf{f})_1 = (\mathbf{D}_1 \mathbf{f})_0,$$

$$f'(\eta_\infty) = 1 \Rightarrow (\mathcal{L}\mathbf{f})_{N-1} = (\mathbf{D}_1 \mathbf{f})_{N-1}.$$

These constraints are a replacement of the respective rows of $\mathcal{L}$, which results in a square, sparse and non-singular operator.

### 3.2.4 Numerical Inversion of the Linear Operator

The inverse operator $\mathcal{L}^{-1}$ needed in the HAM recursion is implemented numerically by solving
$$\mathcal{L}\mathbf{u} = \mathbf{b},$$
Here $\mathbf{b}$ is a known discrete residual vector.
LU factorization of $\mathcal{L}$ is computed which allows efficient evaluation of
$$\mathbf{u} = \mathcal{L}^{-1}\mathbf{b}$$
at each homotopy order, boundary entries of $\mathbf{b}$ are set to zero to maintain the constraints imposed.

### 3.2.5 Discrete Nonlinear Residual

For a given approximation $\mathbf{F}$, the nonlinear Blasius residual is evaluated pointwise as
$$\mathbf{R} = \mathbf{D}_3 \mathbf{F} + \frac{1}{2} \mathbf{F} \odot (\mathbf{D}_2 \mathbf{F}),$$

where $\odot$ denotes element-wise multiplication. This formulation preserves the full nonlinear structure of the governing equation within the discrete HAM framework.
The proposed discretization allows HAM to be converted into a scalable and computationally efficient framework by substituting analytical differentiation and symbolic inversion with sparse numerical operators, and preserves the rigorous homotopy structure of HAM. This numerical basis is necessary in the further incorporation of optimization and physics-informed learning into PI-OHAM.

## 3.3 Recursive HAM Series Computation

This section provides the recursive building of the homotopy analysis series in the proposed numerical PI-OHAM framework. The formulation is a closely follows the philosophy of classical HAM and allows a completely discrete and computationally efficient implementation to nonlinear differential equations.

### 3.3.1 Zeroth-Order Approximation

Let $f(\eta)$ denote the unknown solution of the nonlinear boundary-value problem. The HAM solution is represented as an infinite series

$$f(\eta) = f_0(\eta) + \sum_{m=1}^{\infty} f_m(\eta),$$

where $f_0(\eta)$ is a preliminary approximation to the solution that satisfies the boundary conditions.
Under the current model, a parametric initial guess is chosen as.

$$f_0(\eta; a) = \eta - \frac{1 - e^{-a}}{a},$$

where $a > 0$ is an auxiliary shape parameter optimized alongside the convergence-control parameter $\hbar$. This form satisfies

$$f_0(0) = 0, f_0'(0) = 0, \lim_{\eta \to \infty} f_0'(\eta) = 1.$$

The discrete form of its representation on the computational grid is represented by the vector.

$$\mathbf{f}_0 = [f_0(\eta_0), f_0(\eta_1), \dots, f_0(\eta_{N-1})]^T.$$

### 3.3.2 Zeroth-Order Deformation Equation

After HAM a homotopy between the first guess and the actual solution is built with the help of the embedding parameter $p$ in [0,1]:

$$(1 - p)\mathcal{L}[\Phi(\eta; p) - f_0(\eta)] = p\,\hbar\,\mathcal{N}[\Phi(\eta; p)],$$

where $\mathcal{L}$ is the auxiliary linear operator and $\mathcal{N}$ denotes the nonlinear operator

$$\mathcal{N}[f] = f''' + \frac{1}{2}ff''.$$

The solution $\Phi(\eta; p)$ is assumed to be analytic in $p$ and expanded as

$$\Phi(\eta; p) = f_0(\eta) + \sum_{m=1}^{\infty} f_m(\eta)p^m.$$

### 3.3.3 m-th Order Deformation Equation

Differentiating the zeroth-order deformation equation $m$ times with respect to $p$, dividing by $m!$, and setting $p = 0$, the following recursive linear problem is obtained:

$$\mathcal{L}[f_m(\eta) - \chi_m f_{m-1}(\eta)] = \hbar\,R_m(\eta), m \geq 1,$$

where

$$\chi_m = \begin{cases} 0, & m \leq 1, \\ 1, & m \geq 2, \end{cases}$$

and $R_m(\eta)$ is the $m$-th order HAM residual.

### 3.3.4 Residual Construction

In the case of the Blasius equation, the residual $R_m$ is defined as
$$R_m(\eta) = f'''_{m-1}(\eta) + \frac{1}{2}\sum_{k=0}^{m-1} f_k(\eta)\, f''_{m-1-k}(\eta).$$

Let fm denote the vector representation of fm (η) in discrete form. The remaining vector is calculated as
let $\mathbf{f}_m$ denote the vector representation of $f_m(\eta)$ in discrete form. The residual vector is computed as
$$\mathbf{R}_m = \mathbf{D}_3 \mathbf{f}_{m-1} + \frac{1}{2}\sum_{k=0}^{m-1} \mathbf{f}_k \odot (\mathbf{D}_2 \mathbf{f}_{m-1-k}),$$

where $\odot$ denotes element-wise multiplication. This formulation fully preserves the nonlinear convolution structure of the original equation without linearization or averaging.

### 3.3.5 Recursive Solution of Deformation Equations

The discrete $m$-th order deformation equation is solved as
$$\mathbf{f}_m = \chi_m \mathbf{f}_{m-1} + \hbar\, \mathcal{L}^{-1} \mathbf{R}_m, m \geq 1.$$

$\mathcal{L}^{-1}$ is computed using the sparse LU factorization described in Section 3.2, which is numerically stable and efficient
Each correction term $\mathbf{f}_m$ automatically satisfies the boundary conditions due to their enforcement within $\mathcal{L}$.

### 3.3.6 Series Truncation and Convergence Criterion

Practically, the series is truncated at order $M$ to give the approximate solution.
$$\mathbf{F}^{(M)} = \sum_{m=0}^{M} \mathbf{f}_m.$$

A term-wise stopping criterion is used to make sure that convergence is achieved:
$$\|\mathbf{f}_m - \mathbf{f}_{m-1}\|_\infty < \varepsilon_{\text{HAM}},$$

where $\varepsilon_{\text{HAM}}$ is a prescribed tolerance.
Additionally, convergence is monitored through the decay of the total physics-informed loss defined in Section 3.4.
This recursive numerical HAM formulation does not involve symbolic differentiation or integration, but still has the homotopy structure. Combined with the parameter optimization and physics-informed residual minimization, it is the computational core of the proposed PI-OHAM framework.

### 3.4 Physics-Informed Optimization Strategy

In order to address the convergence sensitivity and manual parameter tuning of classical HAM, the proposed PI-OHAM framework integrates a physics-informed optimization strategy that jointly identifies the optimal convergence-control parameter $\hbar$, the initial-guess shape parameter $a$, and the effective truncation order $M$. The governing equations, boundary conditions and optional data constraints are imposed in a unified manner.

### 3.4.1 Physics-Informed Loss Functional

The physics-informed loss functional is a functional that depends on physics to determine the loss.
Let $F(\eta)$ denote the truncated HAM solution of order $M$,

$$F(\eta) = \sum_{m=0}^{M} f_m(\eta),$$

with derivatives computed numerically. A composite physics-informed loss functional is defined as

$$\mathcal{J}(\hbar, a; M) = \mathcal{J}_{res} + \lambda_{bc}\mathcal{J}_{bc} + \lambda_{data}\mathcal{J}_{data},$$

where each component measures a distinct physical constraint.

### 3.4.2 Governing-Equation Residual Loss

The nonlinear differential equation is imposed at all the collocation points by the residual loss:

$$\mathcal{J}_{res} = \frac{1}{N}\sum_{i=1}^{N}\left[F'''(\eta_i) + \frac{1}{2}F(\eta_i)F''(\eta_i)\right]^2.$$

In discrete form,

$$\mathcal{J}_{res} = \frac{1}{N}\parallel \mathbf{D}_3\mathbf{F} + \frac{1}{2}\mathbf{F}\odot(\mathbf{D}_2\mathbf{F})\parallel_2^2.$$

This residual is also computed on the complete reconstruction rather than on averaged or truncated expressions, as in classical HAM, and thus the full nonlinear physics is retained.

### 3.4.3 Boundary-Condition Loss

Boundary conditions are enforced weakly through a penalty formulation:
$$\mathcal{J}_{bc} = F(0)^2 + F'(0)^2 + (F'(\eta_\infty) - 1)^2.$$

This approach guarantees numerical robustness while avoiding over-constraining the linear deformation equations.

### 3.4.4 Data-Consistency Loss (Optional)
In case reference or experimental data $f^{ref}(\eta)$ are known, a data-driven consistency term is added:

$$\mathcal{J}_{data} = \frac{1}{N}\sum_{i=1}^{N}[F(\eta_i) - f^{ref}(\eta_i)]^2.$$

The term allows inverse modeling and hybrid physics-data learning without affecting the analytical structure of HAM.

### 3.4.5 Joint Optimization of $\hbar$ and $a$

The best parameters are determined by solving.
$$(\hbar^*, a^*) = \arg \min_{\hbar, a} \mathcal{J}(\hbar, a; M),$$

with admissible limits,
$$\hbar \in [\hbar_{\min}, \hbar_{\max}], a \in [a_{\min}, a_{\max}].$$

Two stage optimization strategy is used:
1. **Coarse global sweep** over a tensor grid in $(\hbar, a)$ to identify a robust initial basin.
2. **Local gradient-based refinement** using the L-BFGS-B algorithm for rapid convergence.

This hybrid strategy avoids local minima while maintaining computational efficiency.

### 3.4.6 Order-Adaptive Optimization and Convergence Monitoring

The optimization is done in an incremental manner with respect to the HAM of order $M$. For each order:
$$\mathcal{J}_M^* = \min_{\hbar, a} \mathcal{J}(\hbar, a; M).$$

The iteration is stopped when either
$$\mathcal{J}_M^* < \varepsilon_{\text{tol}} \text{ or } |\mathcal{J}_M^* - \mathcal{J}_{M-1}^*| < \varepsilon_{\text{imp}},$$

making sure that the effective truncation order is determined automatically. The process is stopped when the loss improvement falls below a prescribed tolerance or when the total loss reaches machine-level accuracy.

### 3.4.7 Physical Interpretation and Advantages of PI-OHAM

- The optimization automatically chooses the best $\hbar$, removing the analysis of $\hbar$-curves manually.
- The parameter $a$ of the initial guess shape is optimized to the best fit for the nonlinear solution manifold.
- The residual loss incorporates the entire physics of the considered problem, boundary conditions and optional data.
- The resulting HAM solution has a faster convergence, lower truncation order, and greater accuracy than classical HAM.

This physics-informed optimization transforms HAM from a parameter-sensitive analytical technique into a robust, data-aware, and computationally adaptive solver, forming the core innovation of the proposed PI-OHAM framework.

4. **Results Validation, and Computational Efficiency**

This section provides a detailed evaluation of the suggested PI-OHAM framework in the context of the accuracy of the solutions, the ability to match the benchmark results, and the efficiency of the computations. It is a canonical test case based on the Blasius boundary-layer problem, specifically the wall shear parameter $f''(0)$, minimization of the residual, and runtime comparison with Physics-Informed Neural Networks (PINNs).

## 4.1 Accuracy and Convergence of PI-OHAM

The Physics-Informed Optimal Homotopy Analysis Method (PI-OHAM) is proposed and applied to the classical Blasius boundary-layer problem to determine the accuracy, convergence, and computational performance. The main validation measure is the wall shear parameter $f''(0)$, whose benchmark value is 0.332057.

Table 1 provides a comparative analysis of accuracy and computational cost between the classical HAM results provided by Liao (2010) and the current PI-OHAM framework. At lower solution orders ($M$ =4-10), the current method already attains errors of the order $10^{-2}$–$10^{-3}$ in just a few seconds of CPU time, compared to the classical HAM which needs much larger computational time with relatively larger errors. The computational cost of the classical HAM increases exponentially with the solution order, reaching over 3000 seconds at $M$ =20, whereas the current PI-OHAM algorithm has almost linear growth in computational time and reaches $M$ =35 in about 48 seconds. In addition, the current approach has similar or better accuracy, and the errors are always in the range of $10^{-3}$–$10^{-4}$. This comparison shows clearly that PI-OHAM provides the same or even higher accuracy as the classical HAM and consumes almost two orders of magnitude less computational time, which is its advantage in efficiency and scalability.

Table 1 *Table 1: Comparison of Absolute Error in $f''(0)$ and CPU Time for Classical HAM (Liao, 2010) and the Present PI-OHAM Method at Different Homotopy Orders*

| | Liao, 2010 | | Present results | |
|---|---|---|---|---|
| $M$ (order) | |f''(0)-0.332057| | CPU time (s) | |f''(0)-0.332057| | CPU time (s) |
| 4 | 0.44 | 2.23 | 4.79E-02 | 1.987195969 |
| 6 | 0.309 | 8.69 | 2.20E-02 | 2.755696058 |
| 8 | 1.85E-03 | 27.1 | 1.69E-02 | 3.640052319 |
| 10 | 2.61E-03 | 71.8 | 9.28E-03 | 5.195281029 |
| 12 | 1.56E-03 | 171 | 8.57E-03 | 6.974167585 |
| 14 | 9.55E-04 | 382.1 | 2.82E-03 | 9.148191452 |
| 16 | 2.76E-04 | 816.2 | 3.01E-03 | 12.67760921 |
| 18 | 1.62E-04 | 1637.8 | 4.93E-03 | 15.20370579 |
| 20 | 7.05E-05 | 3145.6 | 2.52E-03 | 18.37086654 |
| 24 | | | 6.42E-04 | 24.90078044 |
| 28 | | | 2.19E-03 | 31.75946903 |
| 32 | | | 1.09E-03 | 40.75448489 |
| 35 | | | 6.60E-04 | 47.59698582 |

## 4.2 PI-OHAM vs PINN Computational Efficiency and Speed-Up Analysis

PI-OHAM is directly compared to a standard PINN implementation with the same stopping tolerance ($10^{-5}$) to quantify computational efficiency. The PINN setup uses 4000 collocation points, 100 boundary points, a deep neural network with four hidden layers with 64 neurons per layer, and the Adam optimizer.

*Table 2 Computational Cost Comparison*

| Method | Accuracy Target | CPU Time (s) | Speed-Up vs PINN |
|---|---|---|---|
| PINN | $10^{-5}$ | 1180.80 | 1× |
| PI-OHAM | $10^{-5}$ | 20.51 | ≈ **57.6× faster** |

The PI-OHAM framework is similar in accuracy and is almost two orders of magnitude faster in computation. This high speed-up is due to the semi-analytical HAM architecture and physics-informed optimization, which do not require costly gradient-based training and huge parameter spaces.

5. **Results and Discussions**

The simulations were run on the free version of Google Colab, which offers a multi-core CPU (usually 2-4 cores) with approximately 12GB RAM. Parallel computing (N_JOBS = -1) was implemented in the code developed for PI-OHAM, distributing computations across all available cores. HAM/PI-OHAM terms and parameter combinations were independently evaluated by each core, coarsely sampling 10 values of $\hbar$ and 10 values of $a$, on 801 grid points and up to 60 series orders. This approach efficiently determined the best parameter ranges, minimizing overall computational time, and preserving convergence in the given total tolerances ($\varepsilon_{tol} = 1e - 7$) and homotopy order improvement tolerance ($\varepsilon_{imp} = 1e - 10$), which guaranteed high accuracy and stability of the solution.

In this section, the computational results are provided in a case where the data is not known and the boundary conditions are well defined. In this regard, the weights of the boundary loss and data loss were 1 and 0, respectively. The calculated findings are presented in Figures 1-13. Figures 1-3 indicate that the PI-OHAM results of $f$, $f'$ and $f''$ are identical to those of standard numerical schemes. Figure 4 presents the loss plots of the differential equation, boundary conditions, data loss (in this case, the data was created by standard numerical schemes, but its weight was set to zero) and total loss. As can be seen in the figure, the data loss is high at the beginning of the series, but as the series order increases and the solution approaches convergence, the data loss reduces, which guarantees a reliable final solution. Figure 5 plots the losses on a logarithmic scale, and it can be seen that the losses on the boundary condition decrease more rapidly than the loss on the differential equation. The overall loss is the sum of the differential equation loss and the boundary conditions loss in the present case. The overall loss is of the order of $10^{-7}$, which proves the high precision of the solution.

Figures 6 and 7 show the behavior of the loss as a function of the PI-OHAM solution order and computational time, respectively. Figure 6 indicates that the total loss on a logarithmic scale decreases almost linearly with the PI-OHAM order, which means that the total loss decreases exponentially with the solution order. Figure 7 indicates the overall time of computation needed to obtain the solution to the desired accuracy. The figure shows that the total loss on a logarithmic scale decreases almost linearly with time, which means that the loss decreases exponentially with respect to computational time as well. Figures 8 and 9 show the absolute error $|f''(0) - 0.332057|$ versus the order of solution and the time spent on the computation, respectively. These values show that the absolute error is nearly exponentially decreasing with the PI-OHAM solution order and computation time. Together, figure 6-9 indicates the high rate of convergence and precision of the approach.

Figure 10 demonstrates the computational time and the PI-OHAM solution order. The time taken by each successive order is low up to order 10, and thereafter it is slightly higher but less than 5 seconds in all orders. The number also shows that it takes less than 50 seconds to calculate the solution to the 35th order, which shows that the method is computationally efficient even at higher series orders.

Figures 11-13 show the optimization of the auxiliary shape parameter $a$ and the convergence-control parameter $\hbar$ with respect to the total loss. Figure 11 presents a scatter plot that determines the local optimization region in the $(\hbar, a)$ parameter space. The greater the concentration of points, the greater the area where the total loss is minimized, which means locally optimal parameter values. This area is seen in the upper-right part of Figure 11, indicating that the locally optimal parameters meet $a > 1.1$ and $\hbar > -0.3$ This area is observed in the upper-right part of Figure 11. Figure 12 shows how the auxiliary parameter $a$ is refined with respect to the total loss, showing how the solution converges as $a$ changes. It is clear that the solution converges at values of $a$ near 1.1. Figure 13 shows the convergence behavior in terms of $\hbar$, and it can be seen that the solution converges to $\hbar$ values near -0.2. Based on the computational findings, the optimized values of $\hbar$ and $a$ are obtained as -0.22222 and 1.118889, respectively.

Figure 14 shows how the total loss and its components change with the order of PI-OHAM solution under various weighting strategies between the boundary condition (BC) loss and data loss. The effect of loss weighting on the convergence rate and the effective order needed to reach a stable solution is also very clear.

Figure 14(a) (BC weight = 0.8, Data weight = 0.2) shows that the dominance of the BC loss causes all the loss components to decrease rapidly in the first few orders. The total and individual residual losses decline rapidly and reach almost saturation at comparatively low orders (around order 8-10). Beyond this order, further increases in the series order result in only marginal improvements, indicating early convergence of the PI-OHAM expansion. The loss of the BC is small at the initial orders, which proves the high boundary enforcement and numerical stability at lower truncation orders.

Figure 14(b) (BC weight = 0.5, Data weight = 0.5) shows that a more balanced weighting results in the convergence over a more moderate range of orders. The overall loss continues to decay exponentially, but at a slower rate than in the BC-dominated situation (14(a)). The losses level off at order 12-15, suggesting that a larger truncation order is needed to obtain the same level of accuracy as Figure 14(a). This action indicates the extra effort required to meet both boundary constraints and data consistency at the same time.

Figure 14(c) (BC weight = 0.0, Data weight = 1.0) shows that when explicit boundary enforcement is not used, the convergence is completely biased to data fitting. This means that the overall and residual losses demand a much larger solution order (more than order 20) to stabilize into a regime. The loss curves also show small oscillations at intermediate orders, which also point to a lower stability and slower convergence. Though convergence is eventually obtained at higher orders, the absence of BC weighting leads to higher computational cost and poor physical consistency.

In general, the findings indicate that the higher the weight of BC loss, the lower the effective PI-OHAM solution order needed to converge, and the higher the weight of data loss, the higher the order of convergence. This shows the important interaction between loss weighting and series truncation order in the control of convergence speed, numerical stability, and computational efficiency of the PI-OHAM method.

Figures 15(a)-15(c) show the distribution of the optimized convergence-control parameter $\hbar$ and auxiliary shape parameter $a$ obtained by minimizing the total loss with various weight combinations of the loss of boundary condition (BC) and data loss, which are the same cases as in Figure 14.

Figure 15(a) in which the BC loss is dominant (BC weight = 0.8, Data weight = 0.2) shows that the optimum parameter values are concentrated in a relatively small region with $a \gtrsim 1.1$ and $-0.35 < \hbar < -0.25$. This concentrated distribution indicates strong constraint enforcement from the boundary conditions, leading to rapid convergence at relatively lower PI-OHAM orders, as is also the case in Figure 14(a). The tighter clustering indicates the increased stability and strength of the solution in the case when the boundary information prevails in the loss function.

In Figure 15(b) when the weight of both BC and data losses (BC weight = 0.5, Data weight = 0.5) are equal, the scatter of optimal points is moderately wider. Though the optimum area is still concentrated around $a$ =1.1 and $\hbar = -0.33$, the dispersion is higher than in Figure 15(a). This is in line with Figure 14(b) in which a larger PI-OHAM order is needed to achieve similar loss levels. The equal contribution of BC and data losses adds more flexibility in the choice of parameters, leading to slower convergence and more sensitivity to the homotopy parameters.

Figure 15(c) shows that the optimal parameter distribution is the most scattered in the case where the solution is driven by data loss only (BC weight = 0.0, Data weight = 1.0). Despite the fact that there is still a dominant cluster around $a = 1.1$ and $\hbar = -0.25$, there is a large dispersion in both parameters. This greater dispersion is associated with Figure 14(c) where convergence is more sensitive to solution orders and has greater oscillations in loss components. The lack of boundary constraints decreases the regularizing effect of the BC loss, which results in slower convergence and more variability in the optimized homotopy parameters.

Generally, Figures 15(a)-15(c) clearly show that the larger the data-loss weight, the larger the optimal parameter region and the higher the order of PI-OHAM solution required, which is in line with the findings of Figure 14. The findings indicate that the inclusion of the information of the boundary conditions does not only speed up the convergence but also stabilizes the optimization of $\hbar$ and $a$. These results also confirm the consistency, stability, and efficiency of the PI-OHAM framework with various weighting strategies.

Figures 16(a)-16(c) indicate how the CPU time per order and the total computational time change with PI-OHAM solution of various loss weight strategies. The per-order CPU time is small and constant in all cases, and it does not exceed approximately 5 s. Nevertheless, the total time to converged solution is less in Fig. 16(a) (around 60-65 s) than in Fig. 16(b) (around 70-75 s) and 16(c) (around 120-125 s). This implies that the loss-weighting approach that has a greater loss of BC and less loss of data converges faster and has better computational efficiency. However, the convergent solutions can be attained even without the presence of BCs, through the availability of data.

Figures 17(a)-17(c) compare the solutions of the standard numerical method, PI-OHAM, and PINN with BC loss weight = 1.0, data loss weight = 0.0, and error tolerance = $10^{-5}$. The findings indicate that the PI-OHAM solution coincides with the standard numerical solution nearly at the whole domain. Conversely, the PINN solution is not entirely consistent with the numerical findings and has significant differences in the middle of the similarity variable $\eta$. Furthermore, the overall computational time of the PINN solution is about 60 times more than PI-OHAM, which is a clear indication of the high accuracy and computational efficiency of the PI-OHAM framework.

**PI-OHAM Grid Independence, Convergence, Stability, and Consistency**

**Grid Independence**

Calculations were done with grid points of 201, 401 and 801. The calculated results exhibited the same convergence behavior with insignificant variations in the final results. Smaller grids gave smoother residuals and only slight gains in final losses were realized after 801 grid points, suggesting that 801 grid points are adequate to reduce discretization errors.

**Convergence**

The PI-OHAM solution convergence is demonstrated by the monotonic decrease of the total loss and absolute error as the solution order and the computational time increase. The almost linear reduction of the loss in a logarithmic scale (Figures 6 and 7) shows exponential convergence of the method. Also, the absolute error $|f''(0) - 0.332057|$ exponentially decreases with the order of the solution and time (Figures 8 and 9), which proves the rapid convergence to the reference solution of the standard numerical schemes.

**Stability**

The PI-OHAM framework is found to be stable by the fact that there is a well-defined and bounded optimization region of the convergence-control parameter $\hbar$ and the auxiliary shape parameter $a$. Figures 11-13 indicate that the overall loss is not very sensitive to minor changes in these parameters in the optimized ranges ($\hbar \approx 0.22$, $a \approx 1.12$). This parameter robustness makes the solution stable throughout an increase in the series order without oscillation or divergence.

**Consistency**

The PI-OHAM solution is consistent by the precise correspondence between the calculated values of f $f$, $f'$, and $f''$ of the solution and the values calculated with standard numerical methods (Figures 13). Moreover, the simultaneous decrease of the differential equation residual and boundary condition losses to the order of $10^{-7}$ is an indication that the governing equations and boundary conditions are met to the required accuracy, and the numerical formulation is consistent.

**Conclusions and future work:**

We proposed PI-OHAM in this work, which is a single framework that connects the classical homotopy analysis with the philosophy of current physics-informed neural network to solve the non-linear differential equations. In contrast to the conventional HAM, the suggested framework removes the manual adjustment of the convergence-control parameter and initial guess auxiliary parameter and incorporates the complete governing physics in the form of residual-based optimization. PI-OHAM provides the same or better accuracy at a significantly lower computational cost than PINNs, and is analytically transparent and interpretable.

Key contributions include:

- A unified PI-OHAM formulation that combines HAM series expansion with physics-informed residual minimization;
- nonlinear PDEs solver that is numerically aided but analytically interpretable;
- better accuracy-time trade-offs were shown on the Blasius boundary-layer problem;
- The PI-OHAM model has exponential convergence, good numerical stability to control parameters, and good consistency to standard numerical solutions. And
- A conceptual bridge between more than thirty years of homotopy-based reasoning and modern physics-inspired machine learning.

These characteristics put PI-OHAM as an effective and robust alternative to purely data-driven PINNs to nonlinear fluid flow and heat transfer problems, especially when interpretability, robustness, and computational efficiency are of paramount importance.

Future research will be aimed at the extension of PI-OHAM to complex, multi-dimensional, and comlpex flow problems to test its strength in problematic situations. The combination of limited experimental or high-fidelity data can also be used to further improve predictive capabilities of systems with partially known or uncertain boundary conditions. The efficiency, stability, and convergence will be enhanced by developing fully adaptive strategies to automatically select convergence-control and auxiliary parameters. Lastly, the framework can be generalized to other nonlinear systems that are governed by differential equations, including magnetohydrodynamics and reaction-diffusion phenomena, and investigated to solve scientific and engineering problems involving PI-OHAM in real-time or at scale.

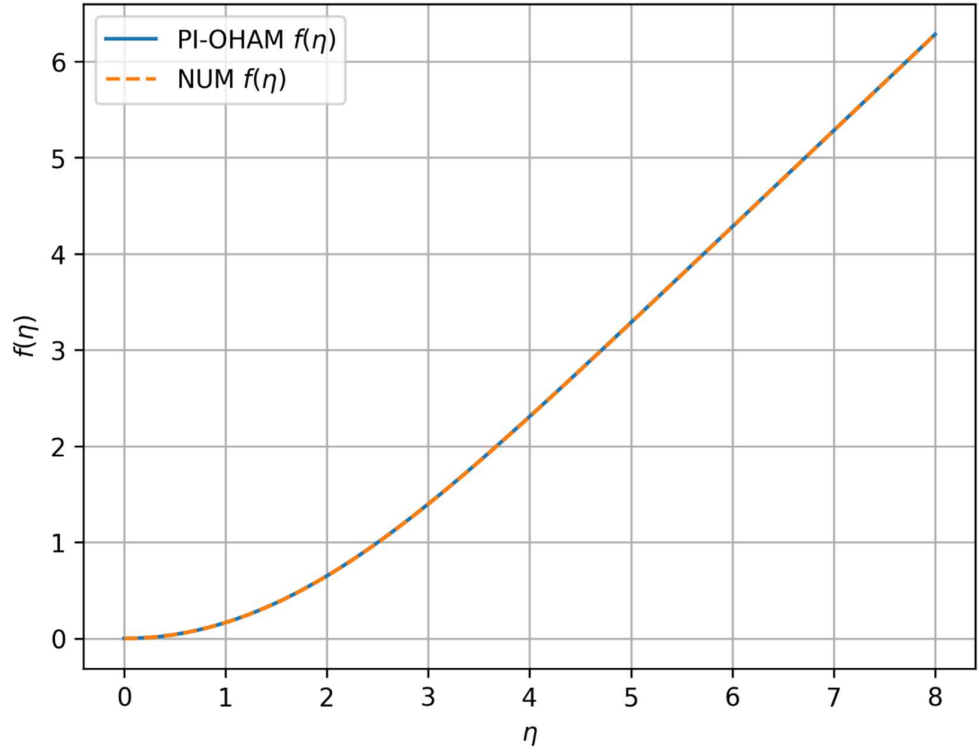

Figure 1

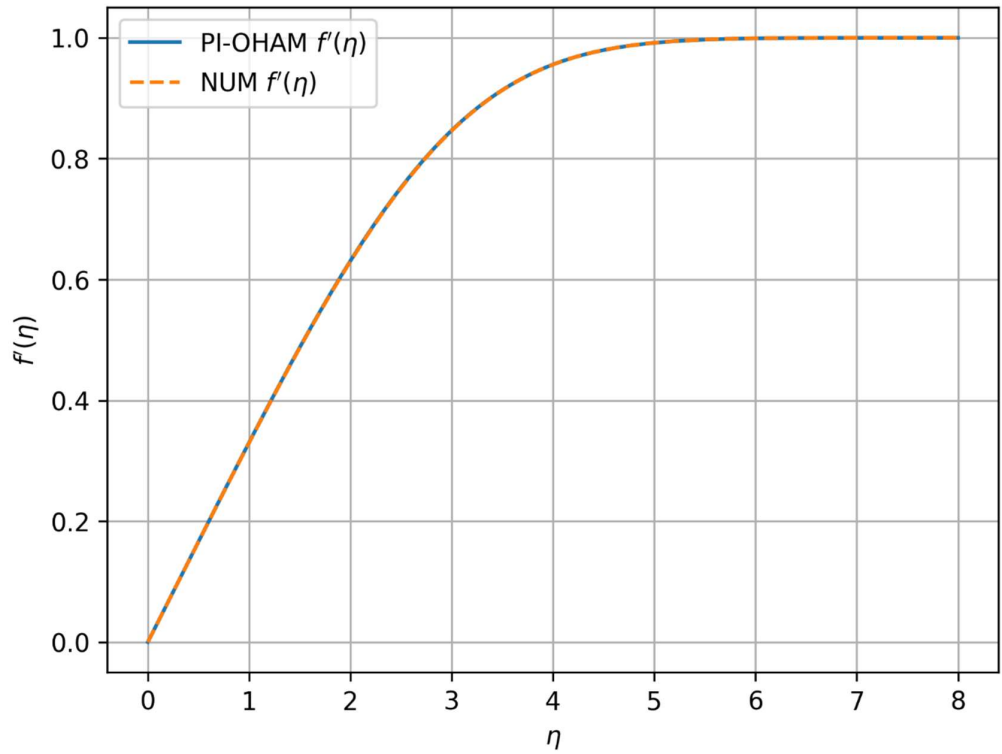

Figure 2

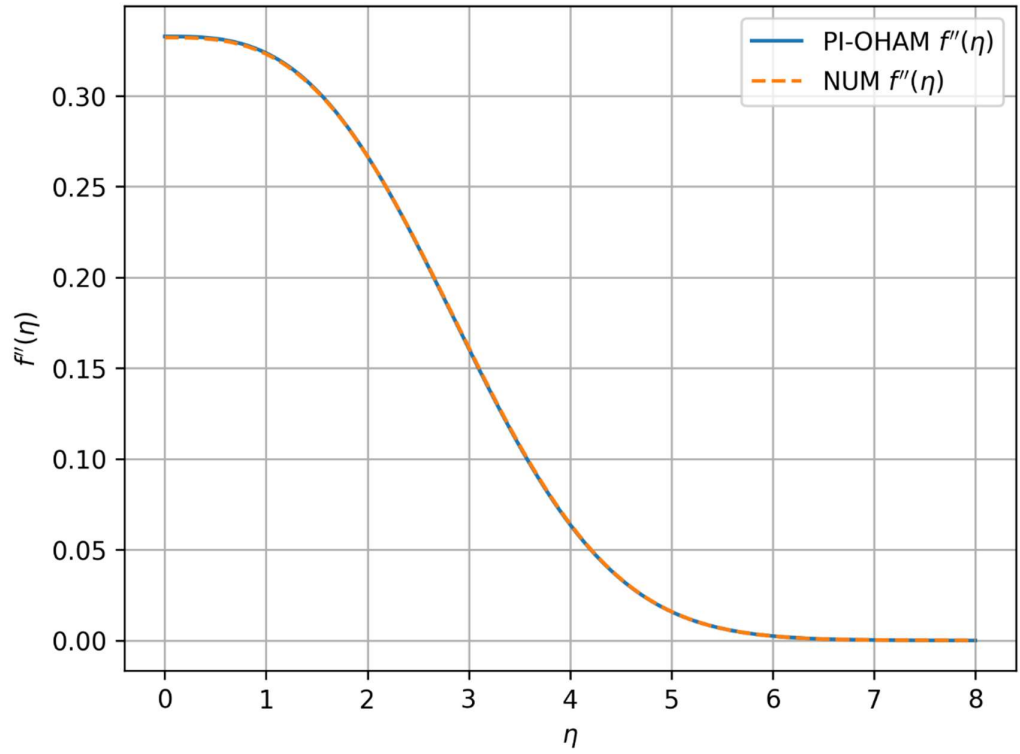

Figure 3

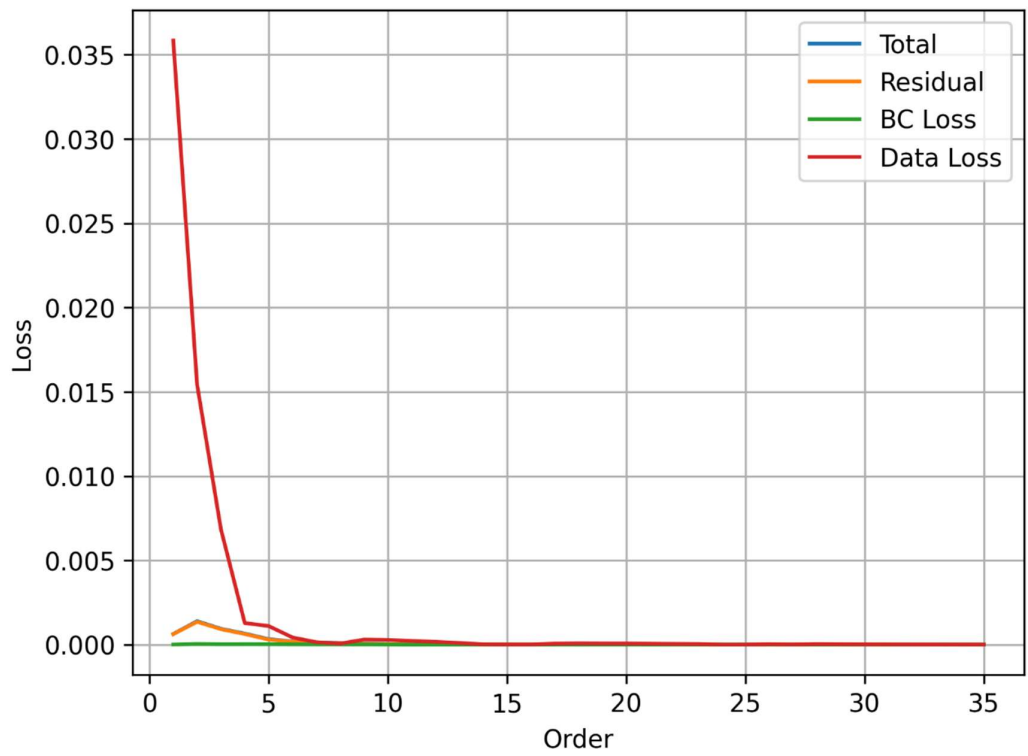

Figure 4

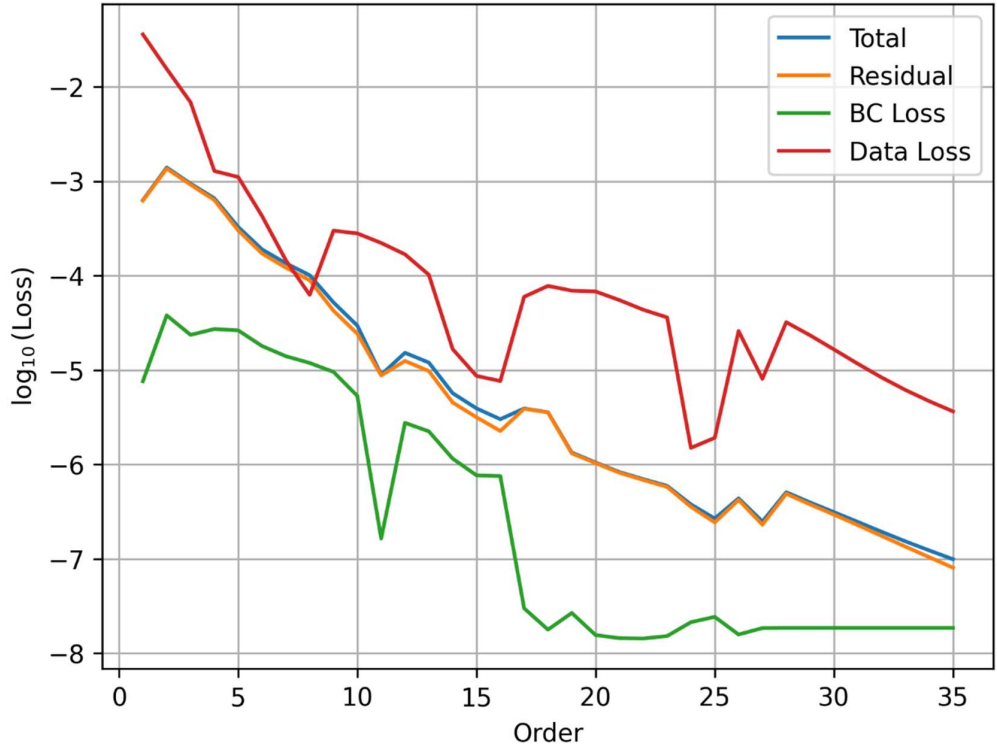

*Figure 5*

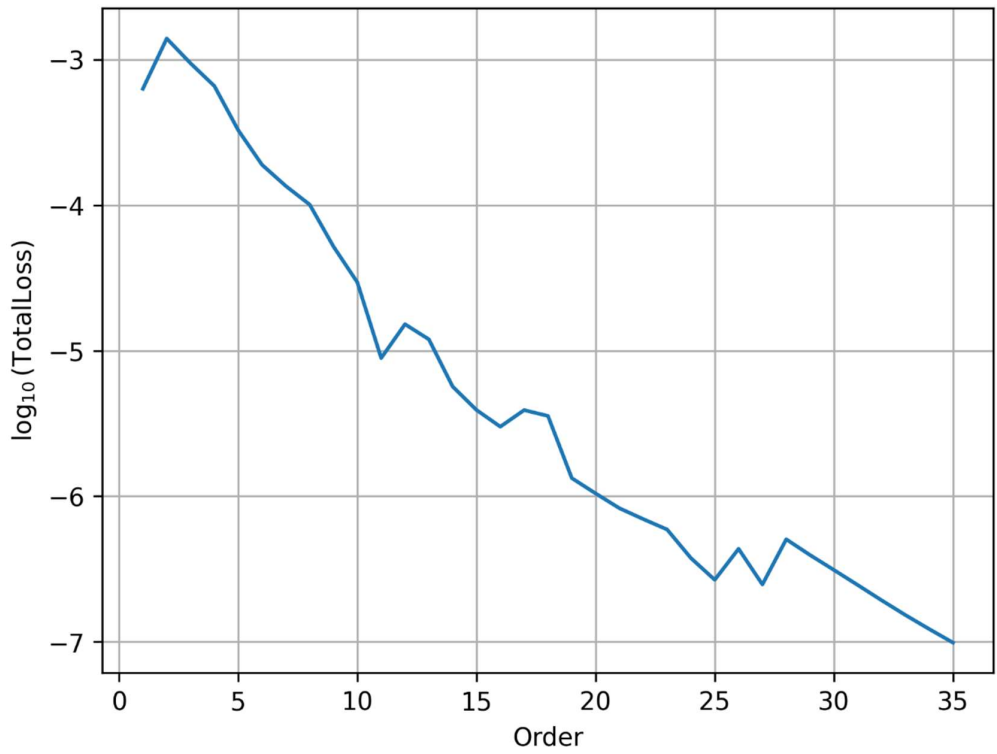

*Figure 6*

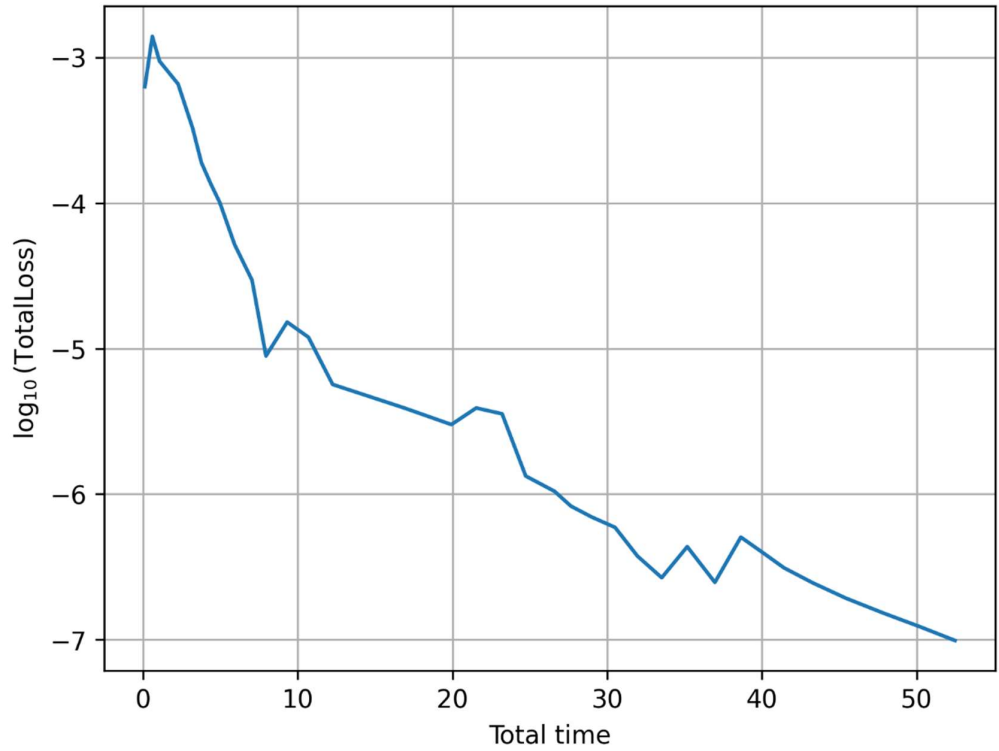

Figure 7

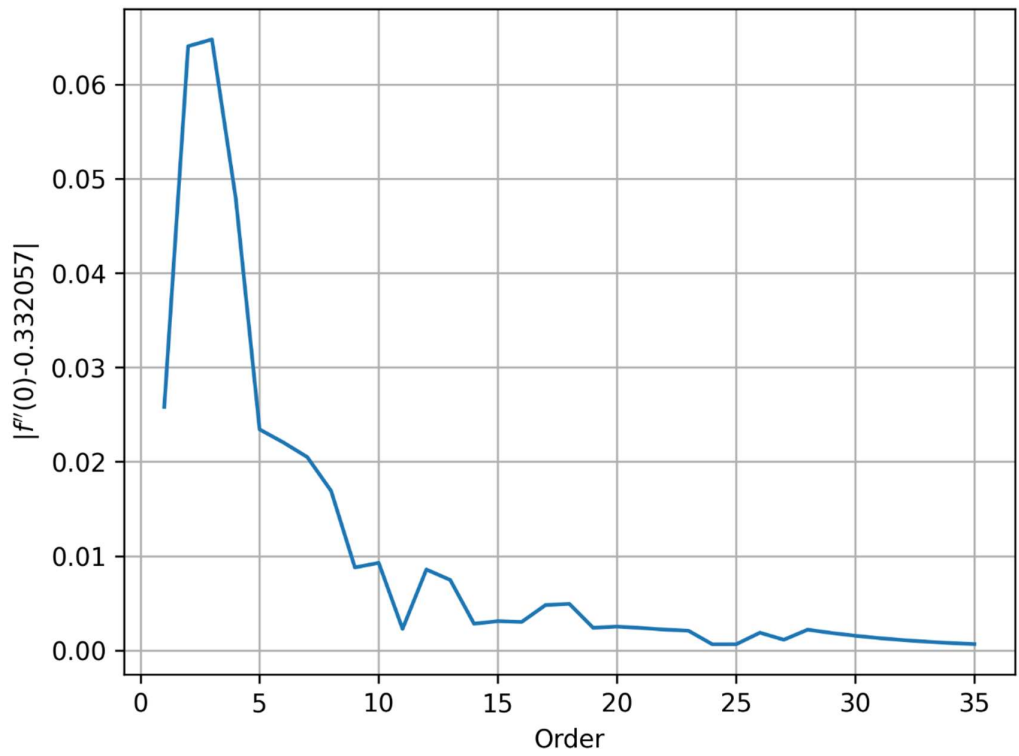

Figure 8

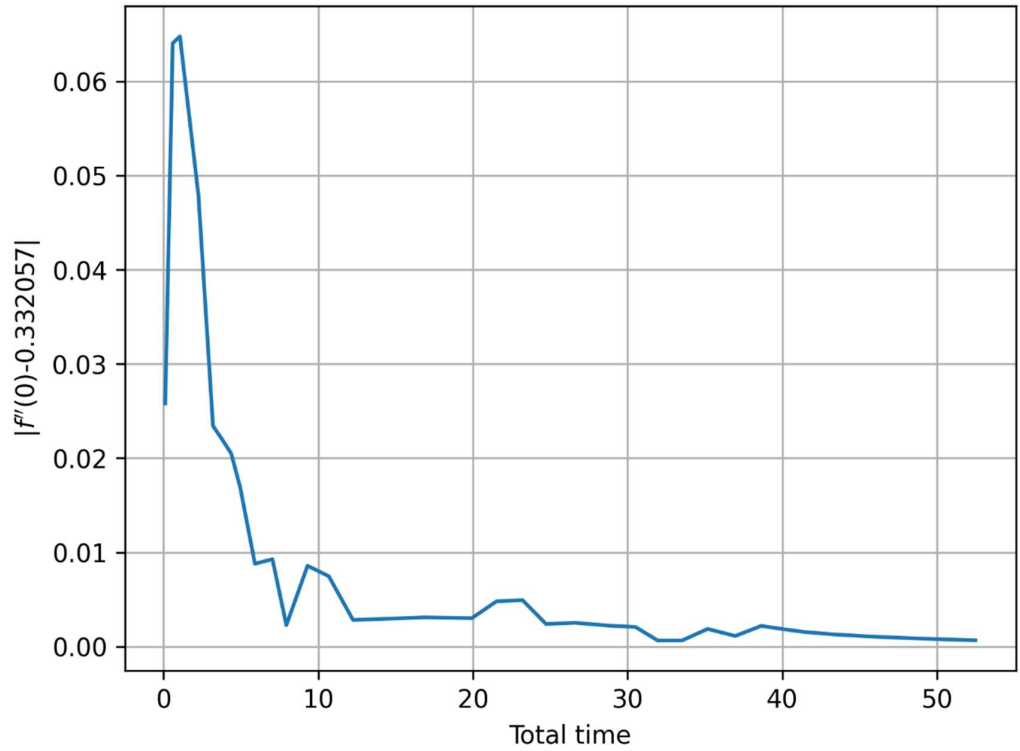

*Figure 9*

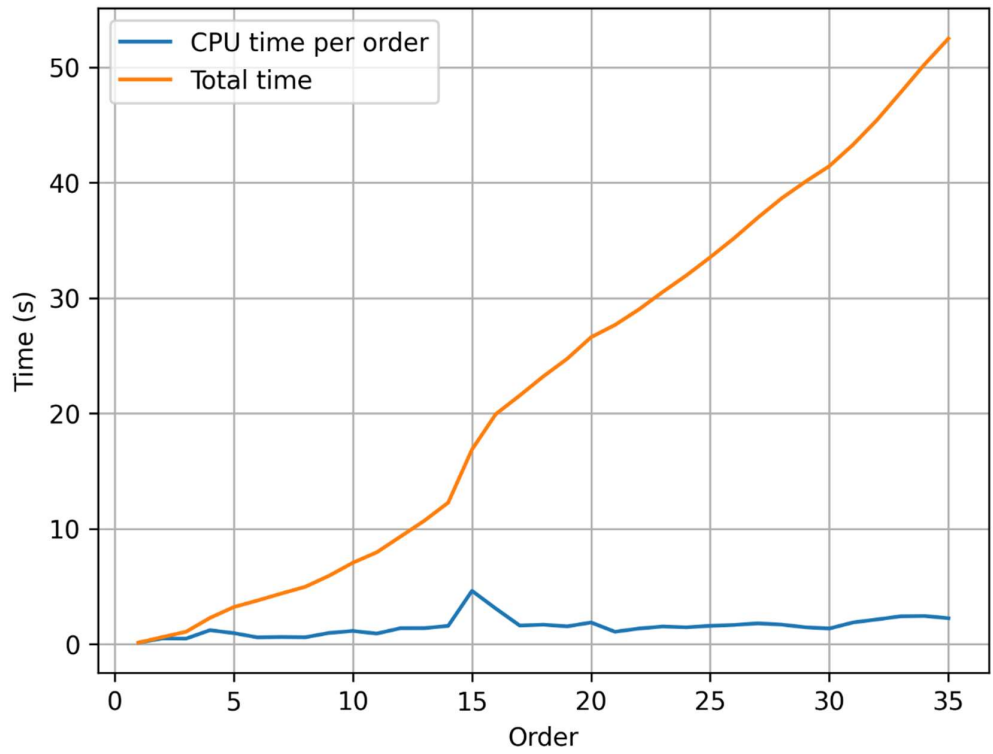

*Figure 10*

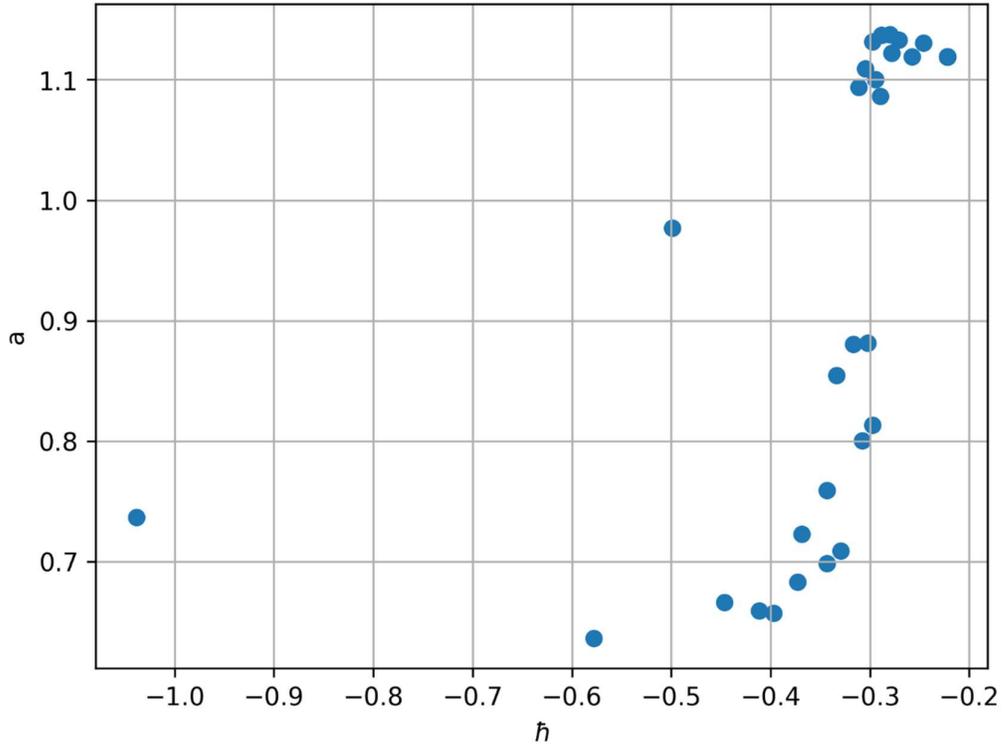

*Figure 11*

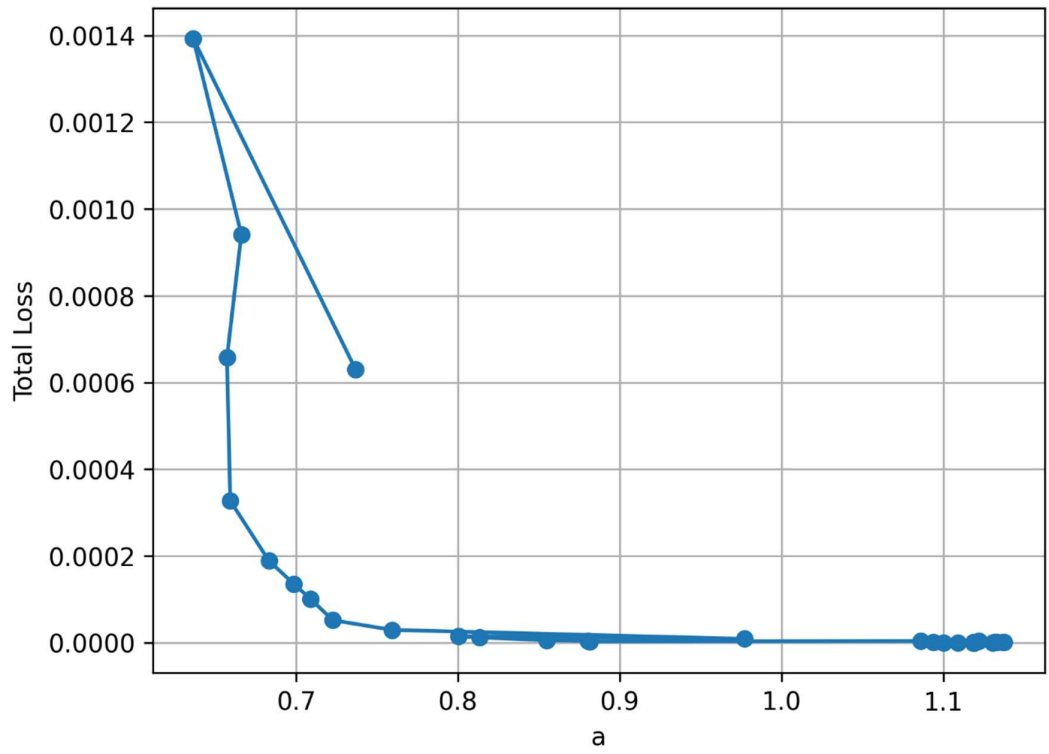

*Figure 12*

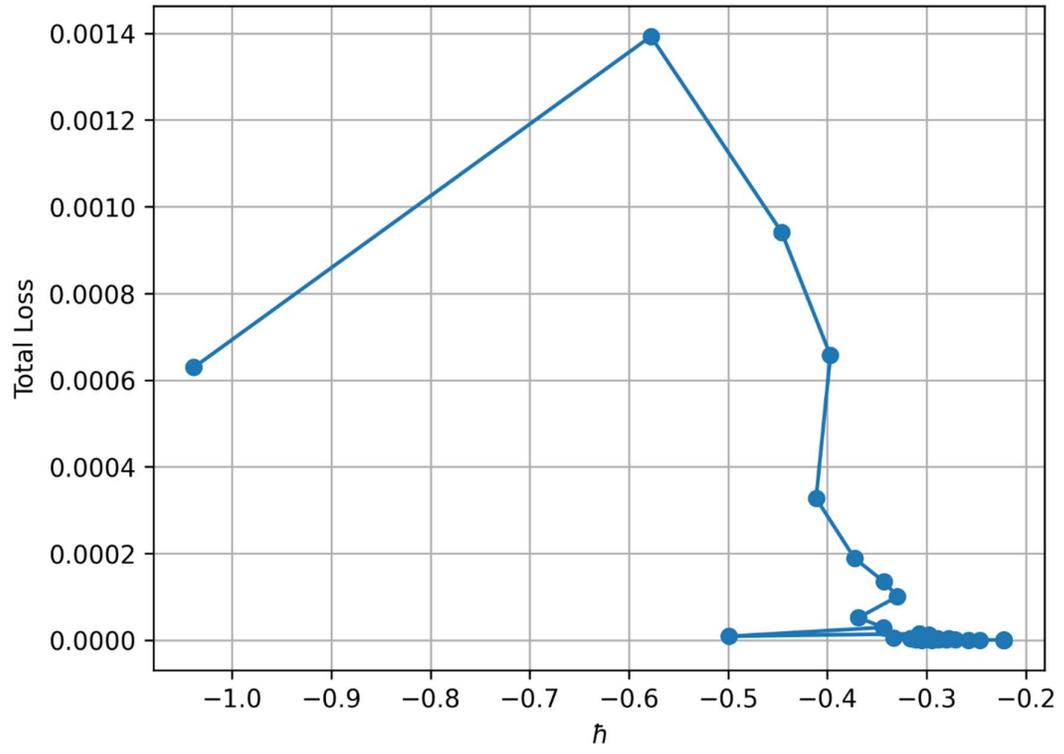

Figure 13

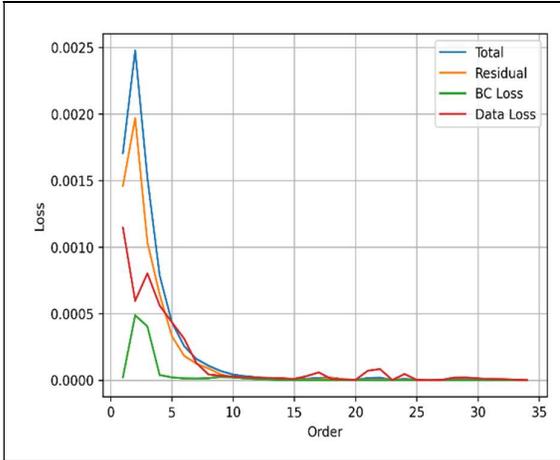
*Figure 14(a) BC weight =0.8, Data weight =0.2*

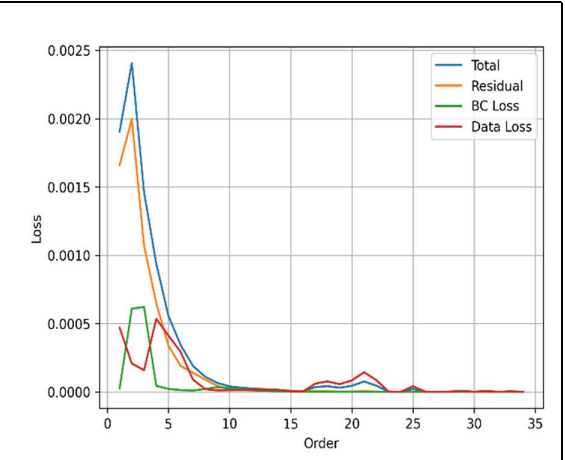
*Figure 14(b) BC weight =0.5, Data weight =0.5*

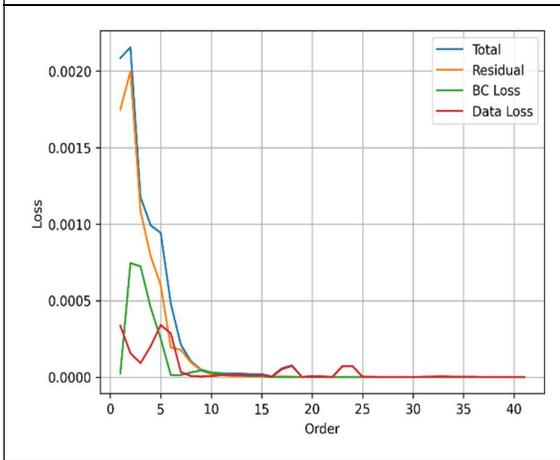
*Figure 14(c) BC weight =0.0, Data weight =1.0*

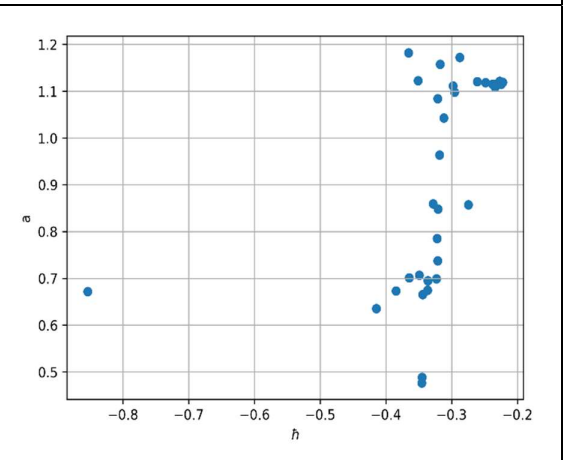
*Figure 15(a) BC weight =0.8, Data weight =0.2*

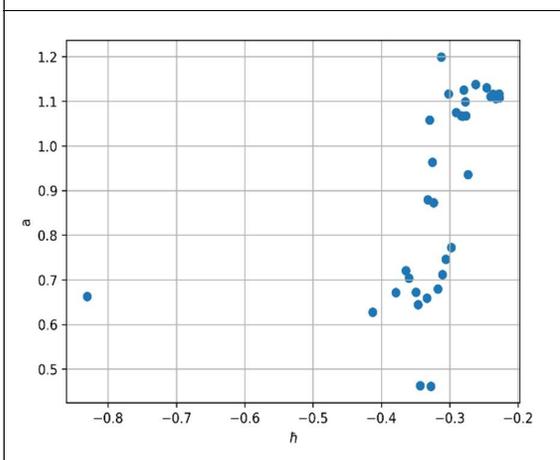
*Figure 15(b) BC weight =0.5, Data weight =0.5*

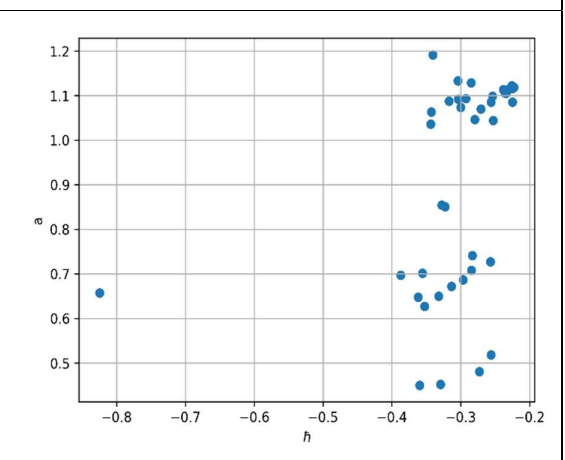
*Figure 15(c) BC weight =0.0, Data weight =1.0*

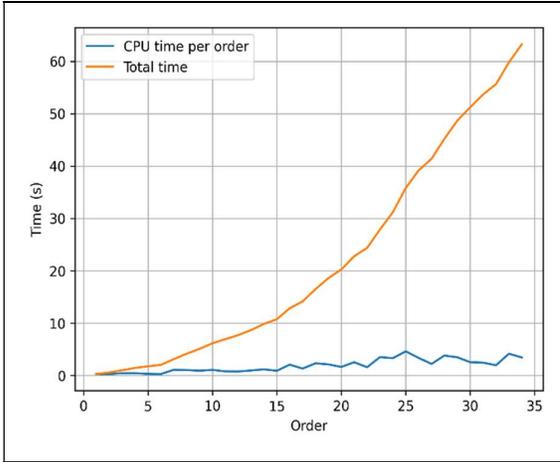

*Figure 16(a) BC weight =0.8, Data weight =0.2*

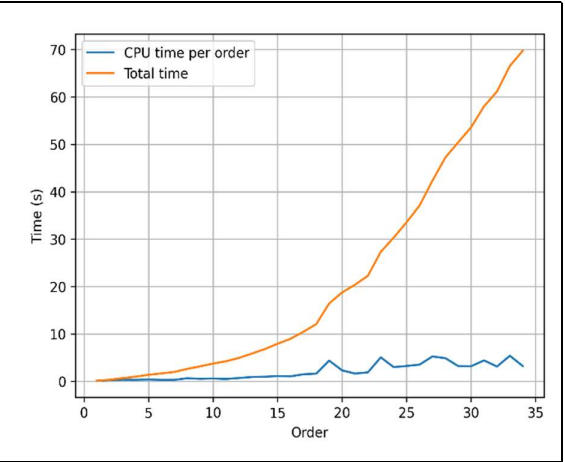

*Figure 16(b) BC weight =0.5, Data weight =0.5*

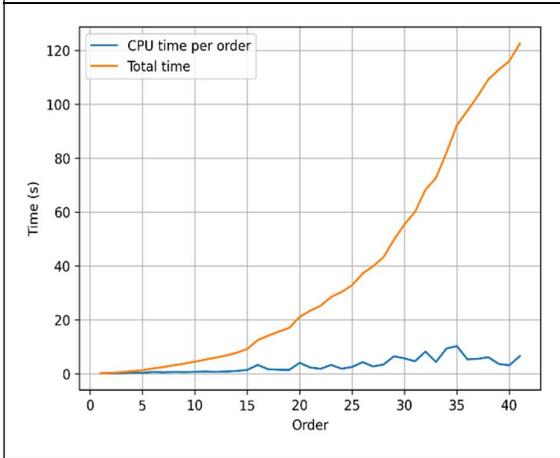

*Figure 16(c) BC weight =0.0, Data weight =1.0*

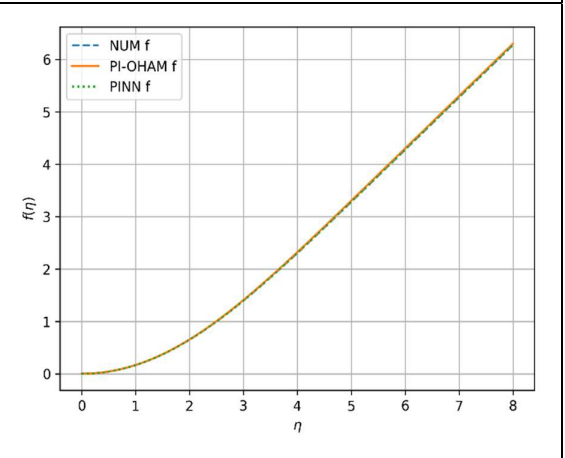

*Figure 17(a) BC weight =1.0, Data weight =0.0, tol=1e-05*

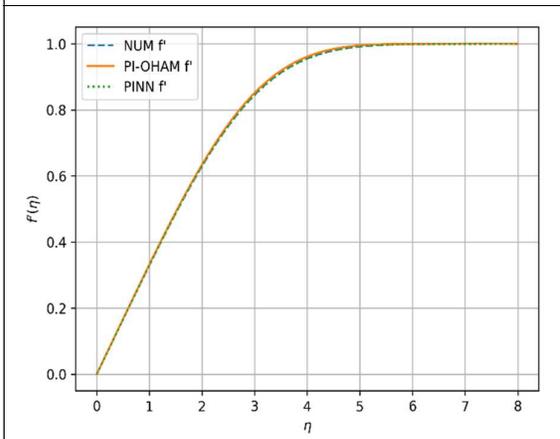

*Figure 17(b) BC weight =1.0, Data weight =0.0, tol=1e-05*

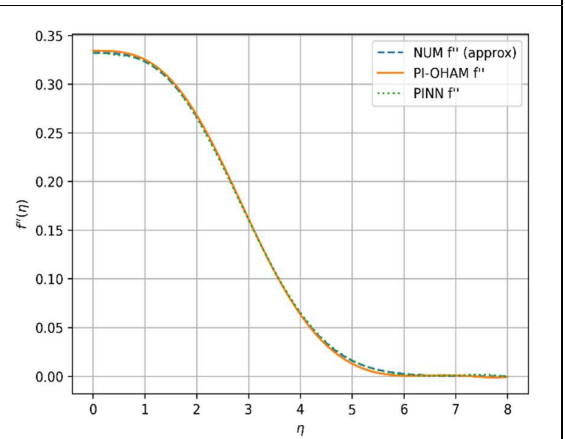

*Figure 17(c) BC weight =1.0, Data weight =0.0, tol=1e-05*